\documentclass[
copyright,
creativecommons]{eptcs}
\providecommand{\event}{EXPRESS/SOS 2014} 

\title{States in Process Calculi}
\author{Christoph Wagner \qquad Uwe Nestmann
\institute{Technische Universit\"at Berlin, Germany}
\email{$\{$christoph.wagner|uwe.nestmann$\}$@tu-berlin.de}}

\def\titlerunning{States in Process Calculi}
\def\authorrunning{C. Wagner, U. Nestmann}

\usepackage[utf8x]{inputenc}
\usepackage[T1]{fontenc}
\usepackage[english]{babel}

\usepackage{amsfonts}
\usepackage{amsmath}
\usepackage{amssymb}
\usepackage{amsthm}
\usepackage{stmaryrd}

\usepackage{tikz}
\usetikzlibrary{arrows,shapes,snakes}

\usepackage{paralist}
\usepackage{booktabs, tabularx}
\usepackage{floatrow}

\usepackage{proof}
\usepackage{algorithmic}
\usepackage{algorithm}
\usepackage{multicol}

\usepackage{commands}
\usepackage{defs}

\usepackage[colorinlistoftodos, disable]{todonotes}

\begin{document}
\maketitle

\begin{abstract}
  Formal reasoning about distributed algorithms (like Consensus) typically requires to analyze global states in a traditional state-based style. This is in contrast to the traditional action-based reasoning of process calculi. Nevertheless, we use domain-specific variants of the latter, as they are convenient modeling languages in which the local code of processes can be programmed explicitly, with the local state information usually managed via parameter lists of process constants. However, domain-specific process calculi are often equipped with (unlabeled) reduction semantics, building upon a rich and convenient notion of structural congruence. Unfortunately, the price for this convenience is that the analysis is cumbersome: the set of reachable states is modulo structural congruence, and the processes' state information is very hard to identify. We extract from congruence classes of reachable states individual state-informative representatives that we supply with a proper formal semantics. As a result, we can now freely switch between the process calculus terms and their representatives, and we can use the stateful representatives to perform assertional reasoning on process calculus models.
\end{abstract}

\section{Introduction}
\label{sec:introduction}

Many articles have been written about the pros and cons of using either action-based (aka: behavioral) or state-based (aka: assertional) formalism for the purpose of verification. This paper can be added to the list of these papers, but here we will not argue in favor of one of the two kinds \cite{denicola90}, and we will also not design a new formalism to integrate both styles \cite{DBLP:conf/acsd/HansenVV03}. We rather deliver support when using an action-based formalism---here: process calculi---to verify the correctness of distributed algorithms, where the reasoning is traditionally rather state-based \cite{lynch:distributed}.  In this context, it is instructive to look at Lamport's case study \cite{lamport82assertional}, where he outlines that assertional reasoning not only applies to sequential and even concurrent algorithms, but also to distributed algorithms \cite{lynch:distributed}. Several of his remarks still hold just like 30 years ago, in particular concerning the requirements and aspects of the assertional method:
\begin{inparaenum}
\item The communication medium needs to be represented explicitly, ideally itself as a process.
\item The concept of a global state should be explicitly available.
\item The method involves reasoning about the entire network of processes at once. 
\end{inparaenum}

We try to profit as much as possible from both worlds.  From process calculi, we take the advantage of exploiting formally defined semantics for a tiny modeling language that is close to a reasonable programming language. From distributed computing, we take the assertional proof methods that are based on the availability of complete global state information to verify the correctness of global invariants. We study the possibility of putting them together such that the assertional proof techniques are explicitly tied to the process calculus code.

A domain-specific process calculus for the particular class of fault-tolerant distributed consensus algorithms should be equipped with 
\begin{inparaenum}[(a)]
\item an explicit notion of distribution, node failure and failure-detection,
\item the underlying communication medium in form of respective primitives,
\item a value language comprising expressions (constructors and selectors) for the required base types and containers (like lists), and
\item some control-flow commands like conditionals.
\end{inparaenum}

Next to the syntax, we also need to supply adequate semantics such as rules to evaluate conditionals, rules to compute the values of expressions, rules to check for the current state of a failure detector so that we know whether we should rather wait a bit longer for a process to react, or just proceed.
For some or most of these enhancements, we can typically choose whether we want their behavior be modeled as explicit steps in a transition system, or whether we hide it in respective rules of a structural congruence relation so that they do not ``cost'' a transition step; we then speak of structural evaluation. 
There are advantages for both.  In essence, as we focus on the interplay between independent distributed components, we most often find the guideline that local computation---which does not change the local state---is hidden within the structural congruence, while interaction with the other components via the interprocess communication or some global failure detection device is modeled as transitions.

With the calculus at hand, one typically uses process constants to define the behavior at the various program points; in a sense, these constants act as labels that we can use to implement recursive behavior. For example, we will later on write an algorithm that starts in the state
\begin{displaymath}
  \prod_{i=1}^{n} \; \loc{i}{\mathsf{B}(v_1,\ldots,v_n)} 
\end{displaymath}
where the notation $\loc{i}{-}$ indicates that the included process description is running at location $i$, with the behavior $\mathsf{B}$ being a constant that is here invoked with its initial set of parameters $(v_1,\ldots,v_n)$.
This hints at a crucial underlying idea: \emph{The modeling philosophy, when using a process calculus for this type of algorithm, is such that we simulate the local state of a process via the list of parameters when invoking a constant that captures its behavior during the intended phase of the algorithm.}  At a later phase, process $i$ is going to be reinstated in a different phase, with a different list of parameters, for example $\mathsf{B} (v'_1,\ldots,v'_n)$.

This simulation of local state information becomes problematic deriving its transition system. The substitution of a process constant by its defining expression (just like a procedure call) takes place within the laws of a structural congruence relation $\equiv$ as part of the rule
\begin{displaymath}
  \ninfer{}
  {\configuration \equiv \configuration_1 \redu{} \configuration_2 \equiv \configuration' }
  {\configuration \redu{} \configuration'}
\end{displaymath}
As a result, the next reachable configuration $\configuration'$ may be in a shape where we do no longer see the parameter lists of local processes.
We might then argue that procedure calls should not be handled by the structural congruence, but rather at the cost of a transition step. However, then we would still lose the structure immediately. We may instead require that after every step, we \emph{must} end up in a form consisting of just process constants. This could be enforced syntactically, or imposed as a modeling discipline.
But then another problem occurs when the behavior of such a process first does some local computation, like evaluating a conditional (without changing its local state!), and depending on this evaluation proceeds with behaviors again conveniently described by other constants, which also do only carry out local computations. Where should the structural evaluation stop? 

Our approach to tackle this dilemma is based on a guideline to identify process constants that are essential to understand the local state of a process. This requires an intuitively deep understanding of the algorithm, just like when one is looking for invariants. Referring to the character of processes as reactive components (sometimes also called event handlers), we should identify the moments when processes wait for messages to arrive. Thus structural evaluation should proceed to the point where such a moment is reached and then stop with the process constant, where such a message arrival is immediately expected. A similar moment of waiting is reached when we require to detect a failure before computation may proceed.

Using this observation, we then design a \emph{standard form} for the global state of the algorithm, in which all and only the possible ``waiting states'' are listed by means of their respective process constants---including their explicit lists of parameters representing the local state information---together with the parallel composition of messages that are underway, i.e.\ sent but not yet received, possibly listed according to their type.  So, a standard form for an algorithm with two possible waiting states ($\mathsf{W}_1,\mathsf{W}_2$) per process may be, as a first approximation, represented as
\begin{displaymath}
  \left(\; 
    \prod_{i\in{I_1}} \mathsf{W}_1 (v^1_1,\ldots,v^1_n)
    \mid
    \prod_{i\in{I_2}} \mathsf{W}_2 (v^2_1,\ldots,v^2_n)
    \;\right)
  \mid 
  M
\end{displaymath}
with $\{I_1,I_2\}$ being a partition of the set of process indices with $I_k$ containing the indices of those processes that are currently in state $\mathsf{W}_k$, and $M$ being the composition of current messages. To make such forms unique, we may have to impose an order on the listing of the various components and message types. Note that this idea is, in a very simple form, already present in Milner's Scheduler example, as found in \cite{Milner:pi-calc}.

We could now try to recover a standard form after each computation step, modulo structural congruence.  (Milner preferred to work with transitions up to strong bisimilarity.)
Instead of trying to guide structural evaluation in order to precisely hit such standard forms, we prefer to extract all the parameters of such standard forms and cast them into a new mathematical structure, basically a sizable tuple that we then use as our substitute representation of the global state.  We devise a mapping that extracts exactly one representative of such a structure from each structural congruence class that is reachable when running the algorithm. The expansion of these representatives back to a process term will then precisely provide the intended standard forms.
As the representatives in the new structure carry the complete state information, we define an explicit operational semantics that produces a new representative after each transition, precisely mimicking the transitions of the expanded process term.  In fact, we design it as a 1-1-correspondence, just like a strong bisimulation between process terms and their stateful representatives.
The goal is then clear.  For any assertional proof of properties of the algorithm, we use the stateful representatives instead of the process terms from which we initially extracted them.

\paragraph*{Contributions}
\label{sec:contributions}

Previously \cite{diplomarbeitvorlage}, we presented a rather detailed analysis of a Consensus algorithm known from the literature \cite{DBLP:journals/jacm/ChandraHT96}. There (and in the related Phd thesis by K\"uhnrich \cite{mortenPhD}), we already introduced the idea of standard forms, but they were not uniquely defined, and we did not design an explicit operational semantics for them. Instead, we focused on the overall proof methodology that integrated the assertional invariant-based reasoning with the definition of a bisimulation relation for the overall correctness proof.  In contrast, in this paper, while we revisit the same algorithm, the new contributions are the precise definition of the standard forms, their uniqueness, the definition of an explicit operational semantics for their stateful representatives, and the proof of 1-1-correspondence with the underlying process calculus semantics.  In the Appendix, we supply some proofs on the correctness of the chosen algorithm, now carried out based on the explicit semantics of the stateful representatives. While the standard forms need to be reinvented for each algorithm, the principles for finding them and also the overall methodology are reusable.



\section{Distributed Consensus}
\label{sec:distr-cons}

In this chapter we introduce a distributed process calculus for fault tolerant systems.
Fault tolerant systems are systems that are able to proceed their work even in the case of failures.
In every fault tolerant system there is a limit to how many failures can occur until the system ceases its operation.
The limit is directly linked to the design of the system and the nature of the failure.
Failures can occur in every part of the system, \EG message transfer, on channels, or processes.
We restrict our interest to permanent crashes of processes.
To detect crashes we need a tool, which we denote as failure detector.
We concentrate on unreliable failure detectors as described in \cite{DBLP:journals/jacm/ChandraT96}.

\subsection{Unreliable Failure Detectors}\label{sec:failure_detectors}

The following ideas and definitions are taken mostly verbatim from \cite{DBLP:journals/jacm/ChandraT96}.

A distributed system consists of a set of $n$ processes, $\Pi = \set{p_1, \ldots, p_n}$.
Every pair of processes is connected by a reliable communication channel.
For simplicity, the existence of a discrete global clock is assumed, with the range $\mathcal{T}$ of the clock's ticks being the set of natural numbers.
The function $F: \mathcal{T} \to 2^{\Pi}$ describes a \emph{failure pattern}, while $F(t)$ denotes the set of processes that have crashed through time $t \in \mathcal{T}$.
The set of crashed processes is described by $crashed(F) = \bigcup_{t \in \mathcal{T}} F(t)$, where the set of correct processes is described by $correct(F) = \Pi - crashed(F)$.
$p \in crashed(F)$ says $p$ crashes in $F$ and $p \in correct(F)$ says $p$ is correct in $F$.
Note, only failure patterns $F$ such that at least one process is correct are considered, \IE $correct(F) \not= \emptyset$.

Each failure detector module outputs the set of processes that it currently suspects to have crashed.
The function $H : \Pi \times \mathcal{T} \to 2^{\Pi}$ is called a \emph{failure detector history}, where $H(p, t)$ denotes the value of the failure detector model of process $p$ at time $t$.
$q \in H(p,t)$ says that $p$ suspects $q$ at time $t$ in $H$.

Informally, a failure detector $\mathcal{D}$ provides (possibly incorrect) information about the failure pattern $F$ that occurs in an execution.
Formally, failure detector $\mathcal{D}$ is a function that maps each failure pattern $F$ to a set of failure detector histories $\mathcal{D}(F)$.
This is the set of all failure detector histories that could occur in executions with failure pattern $F$ and failure detector $\mathcal{D}$.
We use two different failure detectors that satisfy the following completeness and accuracy properties.
\begin{compactitem}
	\item \emph{Strong completeness}: Eventually every process that crashes is permanently suspected by every correct process. \\
		$ \forall F\. \forall H \in \mathcal{D}(F)\. \exists t \in \mathcal{T}\. \forall p \in crashed(F)\. \forall q \in correct(F)\. \forall t' \geq t\. p \in H(q, t') $
	\item \emph{Strong accuracy}: No process is suspected before it crashes. \\ $ \forall F\. \forall H \in \mathcal{D}(F)\. \forall t \in \mathcal{T}, \forall p,q \in \Pi - F(t)\. p \not\in H(q,t) $
	\item \emph{Weak accuracy}: Some correct process is never suspected.\\
		$ \forall F\. \forall H \in \mathcal{D}(F)\. \exists p \in correct(F)\. \forall t \in \mathcal{T}\. \forall q \in \Pi - F(t)\. p \not\in H(q,t) $
\end{compactitem}
A failure detector is called \emph{perfect} if it satisfies strong completeness and strong accuracy, with the set of all these failure detectors being denoted by $\perfect$.
A failure detector is called \emph{strong} if it satisfies strong completeness and weak accuracy, with the set of all these failure detectors being denoted by $\strong$.

\subsection{Syntax of the Calculus}\label{sec:calculus_syntax}

We introduce a tailor-made calculus \cite{diplomarbeitvorlage} to model Distributed Consensus.
It is based on process calculi like the well-known CCS or $\pi$-calculus \cite{CCS, picalculus}.
The calculus is an adapted version of the calculus introduced by Francalanza and Hennessy in \cite{DBLP:conf/esop/FrancalanzaH07}.
Most of the following definitions are taken verbatim from \cite{diplomarbeitvorlage, mortenPhD}.

\begin{table}[tbp]
  \begin{center}
    \begin{tabular}{llll}
      \textsc{Data values} $\values$ & $v$ & ::= & $\bot,0,1,2,3,\ldots~~|~~(v, v)~~|~~\set{v, \ldots, v}$\\[2mm]
      \textsc{Variable pattern} & $X$ & ::= & $x ~~|~~(X, X),  ~~\text{with~} x\in \names$\\[2mm]
      \textsc{Expressions} & $e$ & ::= & $v~~|~~X~~|~~(e,e) ~~|~~\func{f}(e), ~~\text{with~} \func{f} \in \names$\\[2mm]
      \textsc{Guarded processes}~$\guards$ & $G$ & ::= & $\nil ~~|~~\out{c}{e}.P~~|~~\inp{c}{X}.P ~~|~~\susp{k}.P~~|~~\psusp{k}.P ~~|~~ G + G$\\
      & & & $|~~ \myif e \mythen G \myelse G$\\[2mm]
      \textsc{Processes}~$\processes$ & $P,Q$ & ::= & $\tau.P~~|~~G~~|~~K(e)~~|~~P \para P$ \\[2mm]
      \textsc{Networks}~$\networks$ & $M,N$ & ::= & $\nil~~|~~\loc{\ell}{P}~~|~~ N \para N~~|~~N \res a$\\[2mm]
      \textsc{Process equations} & \multicolumn{3}{l}{~~~$D \eqdef \set{K_j(X) = P_j}_{j\in J}$ a finite set of process definitions}
    \end{tabular}
  \end{center}
  \caption{Syntax}
  \label{tab:calculus_syntax}
\end{table}

The syntax of the calculus is shown in \tabref{tab:calculus_syntax}, it consists of four layers:
data values, guarded processes, processes, and networks.
The existence of a countably infinite set of channel, variable, and function names $\textbf{A} = \set{a, b, c, \ldots}$ and a finite set of location names $\locations$ that contains the special name $\star$ is assumed.
The definitions are mostly standard.
$\susp{k}.P$ is the process that behaves like $P$ if the process $k$ is suspected to have crashed and $\psusp{k}.P$ behaves like $P$ if the process $k$ has crashed.
$K(X)$ denotes a parametrized process constant, which is defined with respect to a finite set of process equations $D$ of the form $\set{K_j(X) = P_j}_{j \in J}$.

\begin{definition}[Configurations] Configurations $\configuration$ have
  either of the two forms $\conf{(\L, n)}{\bot}{M}$ or $\conf{(\L,
    n)}{\ti}{M}$, where $\L\subseteq\locations\setminus\set{\wrap}$ is a finite set of
  locations, $n\in\mathbb{N}$ is the number of processes that can crash and $M$ is a network.  
The location $\ti \in   \locations,~\ti \neq \wrap$ is called a \emph{trusted immortal} \cite{DBLP:conf/asian/NestmannF03}; it cannot be suspected and it never crashes. 
We define $\configurations$ as the set of all configurations.
\end{definition}

We define the projection:
\[
\live(\ell,(\L, n)) = \begin{cases}
                       \OP{true} &,\text{if } \ell \in \L \vee \ell = \wrap \\
                       \OP{false} &,\text{else}
                      \end{cases}
\]

Hence $\L \subseteq \P$ is the set of live processes and $n$ denotes the number of processes that are allowed to crash.
Accordingly $\P \res \L$ denotes the set of crashed processes.
Let $\eval{e}$ denote the evaluation of expression $e$, defined in the standard way.

The substitution of value $v$ for a variable pattern $X$ in expression $e$ or process $P$ is
written $e\subst{v}{X}$ and $P\subst{v}{X}$ respectively. The operator
$\fn(\cdot)$ defined on processes and networks is defined as usual. Notice
that only data values can be substituted for names and that all variables
of the pattern $X$ must be free in $P$. We write $\out{c}{e}$ for
$\out{c}{e}.\nil$ and $c.P$ for $\inp{c}{x}.P$, $x\notin\fn(P)$ and $\co{c}$
for $\out{c}{\bot}$.
Moreover we use $ \inpat{a}{x}{i}.P \eqdef \inp{a}{x}.P {\plus} \susp{i}.P\subst{\bot}{x} $, \IE either there is communication on channel $a$ or location $i$ is suspected.
We proceed with $P$ in both cases with respect to the given substitution.

%

\subsection{Semantics of the Calculus}\label{subsec:calculus_semantics}


Network evaluation defines some rules to reduce configurations.

\begin{definition}\label{def:network_eval} Let $>$ be the evaluation relation defined on configurations (assuming $\emph{\live}(\ell,\Gamma)$ everywhere, except explicitly stated otherwise), closed under restriction, parallel composition, 
and the following rules:
\newlength{\hlplengthii}
\settowidth{\hlplength}{$\conf{\Gamma}{\ti}{\loc{\ell}{\myif e \mythen P \myelse Q}}$}
\settowidth{\hlplengthii}{$N$}
\begin{align*}
\parbox{\hlplength}{$\conf{\Gamma}{\ti}{\loc{\ell}{P \para Q}}$} & > \conf{\Gamma}{\ti}{\loc{\ell}{P} \para \loc{\ell}{Q}}  {\tag{E1}\label{eval:e1}} \\
\parbox{\hlplength}{$\conf{\Gamma}{\ti}{\loc{\ell}{\nil}}$} & > \conf{\Gamma}{\ti}{\nil} {\tag{E2}\label{eval:e2}}\\
\parbox{\hlplength}{$\conf{\Gamma}{\ti}{\loc{\ell}{P}}$} & > \conf{\Gamma}{\ti}{\nil}, ~~\neg\OP{live}(\ell, \Gamma) {\tag{E3}\label{eval:e4}}\\
\parbox{\hlplength}{$\conf{\Gamma}{\ti}{\parbox{\hlplengthii}{$\nil$} \para N}$} & > \conf{\Gamma}{\ti}{N} {\tag{E4}\label{eval:e5a}}\\
\parbox{\hlplength}{$\conf{\Gamma}{\ti}{N \para \nil}$} & > \conf{\Gamma}{\ti}{N} {\tag{E5}\label{eval:e5b}}\\
\parbox{\hlplength}{$\conf{\Gamma}{\ti}{\loc{\ell}{\out{c}{e}.P}}$} & > \conf{\Gamma}{\ti}{\loc{\ell}{\out{c}{\eval{e}}.P}} \\
\parbox{\hlplength}{$\conf{\Gamma}{\ti}{\loc{\ell}{K(e)}}$} & > \conf{\Gamma}{\ti}{\loc{\ell}{P\subst{\eval{e}}{X}}}, ~~(K(X) \eqdef[] P) \in D\\
\conf{\Gamma}{\ti}{\loc{\ell}{\myif e \mythen P \myelse Q}} & > \conf{\Gamma}{\ti}{\loc{\ell}{P}}, ~~\eval{e} > 0\\
\conf{\Gamma}{\ti}{\loc{\ell}{\myif e \mythen P \myelse Q}} & > \conf{\Gamma}{\ti}{\loc{\ell}{Q}}, ~~\eval{e} = 0.
\end{align*}
Let $C>^*C'$ denote the maximal evaluation of an arbitrary configuration $C$ into $C'$ respecting the rules above, \IE $\forall C,C' \in \configurations\. C >^*C' \eqdef C > \cdots > C' \wedge  \nexists C''\. C' > C''$.
\end{definition}


We took this definition from \cite{diplomarbeitvorlage} and added (similarly to \cite{mortenPhD}) the rules \eqref{eval:e1} to \eqref{eval:e5b} to it.
We removed transitivity to ease the proof of local confluence in Lemma~\ref{lem:confluence}, because otherwise we would not only have to prove diamonds but also triangles.
Note that structural congruence contains the reflexive and transitive closure of $>$.
%
%
%
%
As an example consider $ \conf{\Gamma}{\ti}{\loc{\ell}{\myif 1 \mythen (\out{c}{e}.P \para Q) \myelse R}} > \conf{\Gamma}{\ti}{\loc{\ell}{\out{c}{e}.P \para Q}} $.
Without Rule~\eqref{eval:e1} we cannot evaluate this term any further.
With this Rule we further evaluate $ \conf{\Gamma}{\ti}{\loc{\ell}{\out{c}{e}.P \para Q}} > \conf{\Gamma}{\ti}{\loc{\ell}{\out{c}{e}.P} \para \loc{\ell}{Q}} > \conf{\Gamma}{\ti}{\loc{\ell}{\out{c}{\eval{e}}.P} \para \loc{\ell}{Q}} $.
Intuitively the Rule~\eqref{eval:e1} allows to simplify the syntactic representation of a term as far as possible.
This significantly simplifies the definition of the semantics in Section~\ref{sec:semant-stand-forms}.
Rules \eqref{eval:e2} to \eqref{eval:e5b} remove dead processes from configurations.

Note that in \cite{diplomarbeitvorlage} the Rule~\ref{eval:e1} is added to structural congruence instead of network evaluation.
We find this version to be more intuitive and handy, because it allows us to omit structural congruence in some following proofs.
Note that this decision does not influence the meaning of structural congruence.


\begin{definition}\label{def:equivalence}
 Structural congruence $\equiv\, \subseteq \configurations \times \configurations$ is the least equivalence relation containing $>$, satisfying commutative monoid laws for ($\networks, \Vert, \nil$) closed under restriction and parallel composition.
\end{definition}

Actions $\alpha \in \act$ are of the form $\alpha ::= \tau~|~\outl{c}{v}~|~\inpl{c}{v}$.
The transition relation $\redu{} \subseteq \configurations \times \act \times \configurations$ is the smallest relation satisfying the rules of \tabref{tab:calculus_semantics}.
Rule \R{TI} non-deterministically selects a trusted immortal.
This rule has to be applied initially because all other rules require the trusted immortal set.
\R{Stop} allows agents to cease execution.
\R{Susp} and \R{PSusp} model (perfect) suspicion of agents.
\R{SumL} and \R{SumR} allow to reduce the left or right side of a sum.
Similar \R{Par} and \R{Res} allow to reduce an agent within parallel composition or restriction.
\R{Com}, \R{Tau}, \R{Snd}, and \R{Rcv} model communication, internal steps, sending, and receiving.
Finally, \R{Red} models steps modulo structural congruence.

\newlength{\semanticdistance}

\begin{table}[tbp]
  \begin{center}
 \begin{tabular}{c}
    $\ninfer{\runa{TI}}{\ti \in \L \setminus \set{\star}}{\conf{(\L, n)}{}{M} \redu{\tau} \conf{(\L,n)}{\ti}{M}}$ \hspace{0.5cm}
    $\ninfer{\runa{Stop}}{\ell \neq \ti \land \ell \in \L}{\conf{(\L, n{+}1)}{\ti}{M} \redu{\tau} \conf{(\L{\setminus}\set{\ell}, n)}{\ti}{M} }$\\[3.5mm]
    
    $\ninfer{\runa{PSusp}}{\live(\ell, \Gamma) \land \neg \live(k,\Gamma)}{\conf{\Gamma}{\ti}{\loc{\ell}{\psusp{k}.P}} \redu{\tau} \conf{\Gamma}{\ti}{\loc{\ell}{P}}}$
    \hspace{5pt}
    $\ninfer{\runa{Susp}}{\live(\ell, \Gamma) \land k \neq \ti \land k \neq \ell}{\conf{\Gamma}{\ti}{\loc{\ell}{\susp{k}.P}} \redu{\tau} \conf{\Gamma}{\ti}{\loc{\ell}{P}}}$\\[3.5mm]
    
    $\ninfer{\runa{Tau}}{\live(\ell, \Gamma)}{\conf{\Gamma}{\ti}{\ell[\tau.P]} \redu{\tau} \conf{\Gamma}{\ti}{\ell[P]}}$ \hspace{0.5cm} $\ninfer{\runa{SumL}}{\live(l, \Gamma) \land \conf{\Gamma}{\ti}{\loc{\ell}{G_1}} \redu{\alpha} \conf{\Gamma'}{\ti}{\loc{\ell}{P}}}{\conf{\Gamma}{\ti}{\loc{\ell}{G_1+G_2}} \redu{\alpha} \conf{\Gamma'}{\ti}{\loc{\ell}{P}}}$ \\[3.5mm]

    $\ninfer{\runa{Par}}{\conf{\Gamma}{\ti}{M} \redu{\alpha} \conf{\Gamma'}{\ti}{M'}}{\conf{\Gamma}{\ti}{M \para N} \redu{\alpha} \conf{\Gamma'}{\ti}{M' \para N}}$ \hspace{0.5cm} $\ninfer{\runa{SumR}}{\live(l, \Gamma) \land \conf{\Gamma}{\ti}{\loc{\ell}{G_2}} \redu{\alpha} \conf{\Gamma'}{\ti}{\loc{\ell}{P}}}{\conf{\Gamma}{\ti}{\loc{\ell}{G_1+G_2}} \redu{\alpha} \conf{\Gamma'}{\ti}{\loc{\ell}{P}}}$ \\

    $\ninfer{\runa{Snd}}{\live(\ell, \Gamma)}{\conf{\Gamma}{\ti}{\loc{\ell}{\out{c}{v}}} \redu{\outl{c}{v}} \conf{\Gamma}{\ti}{\nil}}$ ~~~~
    $\ninfer{\runa{Rcv}}{\live(\ell, \Gamma)}{\conf{\Gamma}{\ti}{\loc{\ell}{\inp{c}{X}.P}} \redu{\inpl{c}{v}} \conf{\Gamma}{\ti}{\loc{\ell}{P\subst{v}{X}}}}$\\[3.5mm]

    $\ninfer{\runa{Com}}{\conf{\Gamma}{\ti}{M} \redu{\alpha} \conf{\Gamma}{\ti}{M'}~~~~\conf{\Gamma}{\ti}{N} \redu{\co{\alpha}} \conf{\Gamma}{\ti}{N'}}{\conf{\Gamma}{\ti}{M \para N} \redu{\tau} \conf{\Gamma}{\ti}{M' \para N'}},~~\alpha, \co{\alpha} \neq \tau$\\[3.5mm]
    
    $\ninfer{\runa{Red}}{\configuration \evalcong \configuration_1 \redu{\alpha} \configuration_2 \evalcong \configuration' }{\configuration \redu{\alpha} \configuration'}$~~
    
    $\ninfer{\runa{Res}}{\conf{\Gamma}{\ti}{M} \redu{\alpha} \conf{\Gamma'}{\ti}{M'}}{\conf{\Gamma}{\ti}{M \res a} \redu{\alpha} \conf{\Gamma'}{\ti}{M'\res a}},~~\alpha \neq \outl{a}{v}, \inpl{a}{v}$
  \end{tabular}
 \end{center}
  \caption{Structural Operational Semantics}\label{tab:calculus_semantics}
\end{table}

Let $ \weakredu{} $ denote the reflexive and transitive closure of $ \redu{\tau} $ and let $ \weakredu{\alpha} $ abbreviate $ \weakredu{}\redu{\alpha}\weakredu{} $ if $ \alpha \neq \tau $ and $ \weakredu{} $ else.
Let $\act^*$ be the set of finite sequences of elements in $\act \setminus \set{\tau}$.
We define $\weakredu{\sigma}$ for some sequence $\sigma = \alpha_1, \ldots, \alpha_n$ with $\alpha_{1 \leq i \leq n} \in \act$ as $\weakredu{\alpha_1} \cdots \weakredu{\alpha_n}$ for $n>0$ and else as $\weakredu{}$.
We use the standard definition of weak bisimulation and weak bisimulation up to techniques as described \EG in \cite{sang}.

\subsection{Case Study}\label{sec:case_study}


Distributed Consensus is the following well-known problem: there is a fixed number
$n$ of agents each initially propose a value $v_i$, $1 \leq i \leq n$;
then, eventually, the agents must agree on a common value
$v_i\in\{v_1,\ldots,v_n\}$.
The model in which we study this problem consists of asynchronously
communicating agents that are vulnerable to crash failures.  Each agent
is furthermore equipped with a failure detector,
which can detect whether other agents have stopped or not.
The precise specification of the problem comprises three properties with
temporal logic flavor:
\medskip
\begin{compactdesc}
 \item[Termination:] Every live agent eventually decides some value.
 \item[Agreement:] No two agents decide differently. 
 \item[Validity:] If an agent decides value $v$, then $v$ was proposed by some agent.
\end{compactdesc}
\medskip
Table \ref{algo1} presents an algorithm by Chandra and Toueg
\cite{DBLP:journals/jacm/ChandraT96} that is supposed to solve Distributed
Consensus in the context of failure detector $\strong$. 
%
It is based on three phases.
In the following we give a short description of the algorithm.
The system consists of $n$ agents, which are identified by numbers, and the wrapper, \IE $\P \eqdef \set{\wrap, 1, \ldots, n}$.
Every agent $p$ has a knowledge vector $V_p$
initially only containing its own proposed value $v_p$, \IE $V_p(p) = v_p$.
Unknown values are represented by $\bot$.

\setlength{\columnsep}{2pt} 
\setlength{\columnseprule}{0.5pt} 

\begin{table}[tbp]
    \begin{minipage}{11cm}
      \begin{multicols}{2}
        \begin{algorithmic}[1]
          \STATE {\bf Pseudo code for agent $p$}
          \STATE $\knowvec_p \leftarrow \botvec$,~~$\knowvec_p(p) \leftarrow v_p$
          \STATE $\comvec_p \leftarrow \knowvec_p,~~M_p \leftarrow \emptyset$
          \STATE
          \STATE {\bf Phase 1:}
          \STATE {\bf for all} $r_p \leftarrow 1$ to $n{-}1$ {\bf do}  
          \STATE \TAB \send $\phasei(p, r_p, \comvec_p)$ to all
          \STATE \TAB $\comvec_p \leftarrow \botvec$
          \STATE \TAB \block
          \STATE \TAB \TAB \textbf{for all} $1 \leq q \leq n$ 
          \STATE \TAB \TAB \TAB \receive $m=\phasei(q, r_p, \comvec)$
          \STATE \TAB \TAB \TAB $M_p \leftarrow M_p \cup \set{m}$
          \STATE \TAB \TAB \TAB {\bf or} ~suspect~ $\susp{q}$
          \STATE \TAB  {\bf for all} {$q \leftarrow 1$ to n}  {\bf do} 
          \STATE \TAB \TAB {\bf if~} {$\knowvec_p(q) = \bot$ {\bf and} $\exists \comvec' \in M_p$}
          \STATE \TAB \TAB  with $\comvec'(q) \neq \bot$ {~\bf then}
          \STATE  \TAB \TAB \TAB $\knowvec_p(q) \leftarrow \comvec'(q)$
          \STATE  \TAB \TAB \TAB $\comvec_p(q) \leftarrow \comvec'(q)$
          \STATE
          \STATE {\bf Phase 2:}
          \STATE \send $\phaseii(\knowvec_p)$ to all
          \STATE \block
          \STATE \TAB \textbf{for all} $1 \leq q \leq n$ {\bf do}
          \STATE \TAB \TAB \TAB \receive $m=\phaseii(\knowvec)$
          \STATE \TAB \TAB \TAB $M_p \leftarrow M_p \cup \set{m}$
          \STATE \TAB \TAB \TAB {\bf or} ~suspect~$\susp{q}$
          \STATE  {\bf for all} {$q \leftarrow 1$ to n}  {\bf do} 
          \STATE  \TAB {\bf if~}{$\exists \knowvec' \in M_p: \knowvec'(q) = \bot$}
          \STATE  \TAB \TAB  {\bf then}~$\knowvec_p(q) \leftarrow \bot$
          \STATE
          \STATE {\bf Phase 3:} 
          \STATE $\mbox{decide} = \min{\makeset{q}{\knowvec_p(q) \neq \bot}}$
        \end{algorithmic}
      \end{multicols}
    \end{minipage}
  \caption{Distributed Consensus \cite{DBLP:journals/jacm/ChandraT96}}\label{algo1}
\end{table}

Validity is shown by an invariant.
To show termination and agreement, we have to prove following bisimulation:
\[
  \conf{(\P, n-1)}{\bot}{(\OP{System} \para \OP{Wrapper}) \res R} \approx \conf{(\P, 0)}{\ti}{\loc{\wrap}{\co{ok}}}
\]

\begin{compactdesc}
 \item[Phase~1] consists of $n-1$ rounds.
In every round agent $p$ broadcasts to every other agent a message $\comvec_p$, initially containing the knowledge of the own proposed value.
$p$ then collects all messages $\comvec_q$ from every agent $q$ in the system or suspects $q$ to have crashed, \IE $\susp{q}$.
If a message is received containing the proposed value of agent $q$ which is not known by $p$ it is added to the knowledge of $p$.
After agent $p$ processed all messages, it builds the message to send in the next round by taking all new learned values into this message.
 \item[Phase~2] is used to synchronize knowledge.
Every agent $p$ sends its knowledge vector $V_p$ to every other agent.
Then $p$ collects for every agent $q$ either $V_q$ or suspects $q$.
If $p$ collected a message such that the message contains no knowledge about agent $q$, then $p$ also removes knowledge about this agent, \IE $p$ sets $V_p(q) = \bot$.
 \item[Phase~3.] In Phase~3 every agent decides on a value $v = \min{\makeset{q}{\knowvec_p(q) \neq \bot}}$.
\end{compactdesc}

\setlength{\semanticdistance}{3mm}
\begin{table}[tbp]
\begin{multicols}{2}
\begin{algorithmic}[1]
\STATE $\text{System } \eqdef$\\
\STATE $\TAB \left (\prod_{i=1}^{n} \loc{i}{\mathsf{P1}_i(1, I^0_i, I_i, \emptyset)} \right)$  \\[\semanticdistance] \label{impl:system}
\STATE $\mathsf{P1}_p (r, \knowvec, \comvec, M) \eqdef$ \\
\STATE $\TAB \TAB \myif (r < n) \mythen$ \\ \label{impl:p1_if}
\STATE $\TAB \TAB \TAB \prod_{1 \leq i \leq n} \out{a_{p, i, r}}{\comvec} \para \mathsf{C1}_p (r, \knowvec, M, 1)$\\ \label{impl:p1_send}
\STATE $\TAB \TAB \hspace{-3pt}\myelse$\\
\STATE $\TAB \TAB \TAB \mathsf{P2}_p(\knowvec, M)$\\[\semanticdistance] \label{impl:p1_else}

\STATE $\mathsf{C1}_p (r, \knowvec, M, i) \eqdef$\\ \label{impl:c1_begin}
\STATE $\TAB \TAB  \myif i \leq n \mythen$\\ \label{impl:c1_if}
\STATE $\TAB \TAB \TAB \inpat{a_{i,p,r}}{\comvec}{i}.\mathsf{C1}_p\Big (r, \knowvec, M + (\comvec, r, i), i+1 \Big)$ \\ \label{impl:c1_receive}
\STATE $\TAB \TAB \hspace{-3pt} \myelse$ \\
\STATE $\TAB \TAB \TAB  \mathsf{P1}_p \Big (r+1, \updatek(r, M, \knowvec), $\\ \label{impl:c1_else}
\STATE $\TAB \TAB \TAB \TAB \TAB \TAB \TAB \TAB \TAB \TAB \TAB \updater(r, M, \knowvec), M \Big)$ \\
 
\STATE $\mathsf{P2}_p(\knowvec, M)  \eqdef$\\
\STATE $\TAB \prod_{1 \leq i \leq n} \out{b_{p,i}}{\knowvec}  \para \mathsf{C2}_p(\knowvec, M, 1)$  \\[\semanticdistance]\label{impl:p2}

\STATE $\mathsf{C2}_p (\knowvec, M, i) \eqdef$ \\
\STATE $\TAB \TAB \myif i \leq n \mythen$\\ \label{impl:c2_if}
\STATE $\TAB \TAB \TAB ~\inpat{b_{i,p}}{\knowvec'}{i}.\mathsf{C2}_p(\knowvec, M + (\knowvec', i), i+1)$\\ \label{impl:c2_receive}
\STATE $\TAB \TAB \hspace{-3pt}\myelse$\\
\STATE $\TAB \TAB \TAB \mathsf{P3}_p(\correct(M,\knowvec))$\\[\semanticdistance] \label{impl:c2_else}

\STATE $\mathsf{P3}_p(\knowvec) \eqdef$\\
\STATE $\TAB \out{c_p}{\getfst(\knowvec)}$\\[\semanticdistance] \label{impl:p3}

\STATE $\WRAP(i, v, b) \eqdef$ \\
\STATE $\TAB \myif (b == 1) \mythen$\\ \label{impl:wrapper_if_1}
\STATE $\TAB \TAB \myif (1 \leq i \leq n) \mythen$\\ \label{impl:wrapper_if_2}
\STATE $\TAB \TAB \TAB \psusp{i}.\WRAP(i+1, v, 1)~+$\\ \label{impl:wrap_receive_begin}
\STATE $\TAB \TAB \TAB \inp{c_i}{v'}.$\\
\STATE $\TAB \TAB \TAB \TAB \myif ((v == \bot \wedge v'\, !\!\!= \bot) \vee v == v') \mythen $ \\ \label{impl:wrap_if_3}
\STATE $\TAB \TAB \TAB \TAB \WRAP(i+1, v', 1) \myelse \WRAP(i, v, 0)$\\ \label{impl:wrap_receive_end}\label{impl:wrap_false_knowledge}
\STATE $\TAB \TAB \myelse \myif (i == n+1) \mythen \co{ok}$ \\ \label{impl:wrap_success}
\STATE $\TAB \myelse \WRAP(i, v, 0)$ \label{impl:wrap_stopped}



\end{algorithmic}
\end{multicols}
\caption{Encoding of the Algorithm in \tabref{algo1}}\label{encoding}
\end{table}

We use an adapted version of the implementation of this algorithm as described in \cite{mortenPhD}, see \tabref{encoding}.
The implementation is adapted such that the whole knowledge vector is no longer sent over channel $c$ in Phase~3.
This change does not affect the correctness of the algorithm because this vector is never used by the wrapper.
We also changed the wrapper code.
In the implementation of \cite{mortenPhD}, the wrapper reduced to $\nil$ upon receiving an invalid value $v$.
Invalid has the meaning that the wrapper already learned a value $v$ some agent $p$ decided on and then received a value $v'$ from some other agent $q$ with $V \not= V'$.
Because it might come in handy to know which process decided on that wrong value, we changed the wrapper in a way such that it has an additional parameter which represents whether it can receive messages or not.
Initially the wrapper is allowed to receive messages but sets the parameter to $0$ upon receiving an invalid value.
Observe that $\WRAP(\cdot, \cdot, 0) \approx \nil$.
Since these are the only changes, the adapted version of the implementation has the same behavior as the original version.

\begin{definition}\label{def:agents_and_U}
 Each agent is uniquely identified by a number  $p \in \agents$, where $\agents= \set{1, \ldots, n}$ is the set of agents, and a proposed value $v_p$.
 The tuple $U$ contains all initially proposed values, \IE $U = (v_1, \ldots, v_n)$.
\end{definition}
%
%
 We use $\bot$ for unknown values and define the domain of values $D$ as $\N \cup \set{\bot}$ where $\N = \set{1, 2, 3, \ldots}$ and let $\leq_{nat}$ be the usual ordering relation on the natural numbers.
 The ordering $\leq \subseteq D \times D$ is the least relation containing $\leq_{nat}$ with the additional requirements $\bot \leq \bot$ and $\bot \leq i$ for all $i = 1, \ldots, n$.
 An $n$-vector is a map from set $\set{1, \ldots, n}$ to the set $D$ and $\botvec$ is the $n$-vector $(\bot, \ldots, \bot)$.
 We occasionally regard vectors as ordered lists of values.
 The ordering $\leq$ is extended point-wise to $n$-dimensional vectors and we write $V_1 \leq V_2$ when vector $V_2$ is greater than or equal to $V_1$.
 We say a vector $V$ is valid if $V \leq U$, where $U$ is the vector of Definition~\ref{def:agents_and_U}.
%
The initial vector $V_i^0$ of agent $i$ contains only its proposed value $v_i$ at position $V_i^0(i)$, \IE $V^0_i(i) = (U(i), 0)$ and $V^0_i(j) = (\bot, 0)$ for all $j \in \agents \setminus \set{i}$.
The replay vector is initialized in a similar way, \IE $\comvec^0_i(i) = U(i)$ and $\comvec^0_i(j) = \bot$ for all $j \in \agents \setminus \set{i}$.
%
%
%
The initial configuration is a term in the calculus of the form
$ \conf{(\P, n-1)}{\bot}{\left (\prod_{i=1}^{n} \loc{i}{\mathsf{P1}_i(1, I^0_i, I_i, \emptyset)} \para \loc{\wrap}{\WRAP(1, \bot, 1)} \right)} \res R $
for some $n \in \N$, $ I^0_i(j) = v_i^0 $ if $ j = i $ and $ \bot^0 $ else, and $ I_i(j) = v_i $ if $ j = i $ and $ \bot $ else.
Let $\Po$ denote the set of all initial configurations.
%
To simplify the following consideration we do not regard initial configurations as reachable configurations, but only configurations that are reachable from some initial configuration and in which the trusted immortal is set.
Let $\Pc$ denote the set of reachable terms, \IE $ \Pr = \makeset{\conf{\Gamma}{\ti}{M}}{\exists C \in \Po\. \exists \sigma \in \act^*\. C \weakredu{\sigma} \conf{\Gamma}{\ti}{M}} $.
%

%
%
%
%
%
%


\section{\StandardFormreps}
\label{sec:standard-forms}

A \standardformrep represents the global state of the system.
We observe that the system consists of messages in transit or agents wanting to receive messages.
To describe the state of the system we need variables that represent these messages and agents.
Thus we obtain the sets of all messages of Phase~1, \IE $\Opi$, Phase~2, \IE $\Opii$, and Phase~3, \IE $\Opiii$.
We also get the sets of all agents wanting to receive messages of Phase~1, \IE $\Inpi$, and Phase~2, \IE $\Inpii$.
These sets do not contain a single value like an agent identifier but rather a tuple of related informations, \EG $\Opi$ contains tuples $(p,i,r,\comvec)$.
These tuples provide the information that there is a message from agent $p$ to agent $i$ in Round $r$ containing the value $\comvec$, we also know that this message belongs to Phase~1 because the tuple is contained in $\Opi$.

\begin{definition}\label{def:standard_form}[\StandardFormrep]
 A \normalform of a reachable configuration $C$ is a configuration of the form in \tabref{tab:standard_form} that is structural congruent to $C$, where $\live(p, \sf{\Gamma})$ for all occurrences of $p$.
 The \standardformrep of a \normalform is the vector given by all boxed elements, \IE a \standardformrep has the form:
\[
 \left( \Gamma, \ti, \Opi, \Opii, \Opiii, \Inpi, \Inpii, \Wj, \Ww, \Wb \right)
\]
with
\begin{align*}
 \Opi   &= \makeset{(p,i,r,\comvec)}{(p,i,r) \in \piout \wedge \comvec = \comvec_{p,r}} \\
 \Opii  &= \makeset{(p, i, V)}{(p,i) \in \piiout \wedge V = \Vpii} \\
 \Opiii &= \makeset{(p,v)}{p \in \piiiout \wedge v = \Vpiii} \\
 \Inpi  &= \makeset{(p,r,V,M,i)}{(p,r) \in \picol \wedge V = \Vpi \wedge M = \Mpi \wedge i = \Ipi} \\
 \Inpii &= \makeset{(p, V, M, i)}{p \in \piicol \wedge V = \Vpii \wedge M = \Mpii \wedge i = \Ipii}
\end{align*}
 Let $\Ps$ denote the set of \normalforms.
 Hence $\Pp \supsetneq \Pc \supsetneq \Ps$.
\end{definition}

\begin{table}
\centering
\settowidth{\hlplength}{$\prod_{(p,i,r) \in \sf{\piout}}\,$}
\bigskip
$
\begin{array}{ll}
 \conf{\sf{\Gamma}}{\sf{\ti}}{} \Bigg( & \prod_{(p,i,r) \in \sf{\piout}} \loc{p}{\out{a_{p,i,r}}{\sf{\comvec_{p,r}}}} \para \\
 & \parbox{\hlplength}{$\prod_{(p,i) \in \sf{\piiout}}$} \loc{p}{\out{b_{p,i}}{\sf{\Vpii}}} \para \\
 & \parbox{\hlplength}{$\prod_{p \in \sf{\piiiout}}$} \loc{p}{\out{c_p}{\sf{\Vpiii}}} \para \\
 & \parbox{\hlplength}{$\prod_{(p,r) \in \sf{\picol}}$} \loc{p}{\CI_p\left(r, \sf{\Vpi}, \sf{\Mpi}, \sf{\Ipi} \right)} \para \\
 & \parbox{\hlplength}{$\prod_{p \in \sf{\piicol}}$} \loc{p}{\CII_p\left(\sf{\Vpii}, \sf{\Mpii}, \sf{\Ipii}\right)} \para \\
 & \parbox{\hlplength}{$ $}\hspace{-1pt} \loc{\wrap}{\WRAP(\sf{\Wj}, \sf{\Ww}, \sf{\Wb})} \Bigg) \res R
\end{array}
$

where $\Ipi \leq n, \Ipii \leq n, \Wj \leq n$, $r < n$, and $\Wb \in \set{0,1}$.
\caption{\NormalForm for Reachable Configurations}\label{tab:standard_form}
\end{table}

By Definition~\ref{def:standard_form} every \normalform represents a reachable configuration.
We prove that for each reachable configuration there is a \normalform, and hence a \standardformrep.

\begin{lemma}\label{lem:normalform_for_reachables}
 For every reachable configuration there is a \normalform.
\end{lemma}

We prove this lemma by induction over the number of steps necessary to reach a configuration from an initial configuration.

\begin{definition}[\SFt]\label{def:sf}
Let $\SFt: \Pc → \Ps$ be a function defined by

 \[
  \SF = \renamewrap(\order(\extract(C')))
 \]
where $C >^* C'$, \IE \extract{} is applied to a fully evaluated configuration.

The auxiliary function \extract{} is given by:
\begin{align*}
 \extract(\sconf{N \res R})              &= \sconf{\extract(N) \res R} \\
 \extract(N \para M)                     &= \extract(N) \para \extract(M) \\
 \extract(\loc{\ell}{P})                 &= \loc{\ell}{\translate(P)}
\end{align*}

By Definition~\ref{def:network_eval}, a fully evaluated configuration is a parallel composition of terms of the form as visualized in \tabref{encoding} Lines~\ref{impl:p1_send}, \ref{impl:c1_receive}, \ref{impl:p2}, \ref{impl:c2_receive}, \ref{impl:p3}, \ref{impl:wrap_receive_begin}--\ref{impl:wrap_receive_end}, \ref{impl:wrap_success}, and \ref{impl:wrap_stopped}.
With the function \extract{}, \translate{} is applied to each of these subterms.
For the following terms
\begin{align*}
 Ph1 &\eqdef \inpat{a_{i,p,r}}{\comvec}{i}.\mathsf{C1}_p\Big (r, \knowvec, M + (\comvec, r, i), i+1 \Big)\\
 Ph2 &\eqdef \inpat{b_{i,p}}{\knowvec'}{i}.\mathsf{C2}_p(\knowvec, M + (\knowvec', i), i+1)\\
 W   &\eqdef \psusp{i}.\WRAP(i+1, v)~+\\
     &\TAB \TAB \TAB \inp{c_i}{v'}.\\
     &\TAB \TAB \TAB \TAB \myif ((v == \bot \wedge v'\, !\!\!= \bot) \vee v == v') \mythen\\
     &\TAB \TAB \TAB \TAB \WRAP(i+1, v', 1) \myelse \WRAP(i, v, 0)
\end{align*}
we define \translate{} as:
\begin{align*}
 \translate(Ph1)     &= \CI_p(r, \knowvec, M, i) \\
 \translate(Ph2)     &= \CII_p(\knowvec, M, i) \\
 \translate(W)       &= \WRAP(i, v, 1) \\
 \translate(\co{ok}) &= \WRAP(0, v, 1) \\
 \translate(P)       &= P, \text{ otherwise}
\end{align*}

The function \order{} orders the subterms of the configuration into a \normalform as visualized in \tabref{tab:standard_form}.
First all outputs on channel $a$ are moved to the beginning of the \normalform followed by all outputs on $b$ and $c$.
Then the input guarded terms on channel $a$ and $b$, denoted as $\CI_p(\cdot, \cdot, \cdot, \cdot)$ respectively $\CII_p(\cdot, \cdot, \cdot)$, followed by the Wrapper.
Hence the function \order{} can be implemented using the structural congruence rules for commutativity and associativity of the parallel operator.

We define \renamewrap as:
\begin{align*}
 \renamewrap(C) &= \begin{cases}
                    \sconf{} (M \para \loc{\wrap}{\WRAP(0, \bot, 1)}) \res R & \text{, if } C = \sconf{M \res R} \text{ and} \\
                                                                             & \,\;\, M \text{ does not contain the location } \wrap \\
                    C                                                        & \text{, else}
                   \end{cases}
\end{align*}
If the wrapper was reduced to $\nil$, the function restores the wrapper to obtain a \normalform as visualized in \tabref{tab:standard_form}.
\end{definition}

Note that restriction is moved outwards by the definition above.
Note as well, that there does not exist an inverse function because \SFt is not injective.
However for each configuration $C$ passed to \SFt we can compute an inverse to the \normalform $\SF$ which is structural congruent to $C$.
Therefore we will now define such a function returning a representative for the inverse being denoted by \SFit.

\begin{definition}[\SFit]\label{def:sfi}
Let $\SFit: \Ps → \Pc$ be a function defined by
 \[
  \SFi = \extract'(C)
 \]

The auxiliary function $\extract'$ is given by:
\begin{align*}
 \extract'(\sconf{N \res R})              &= \sconf{\extract'(N) \res R} \\
 \extract'(N \para M)                     &= \extract'(N) \para \extract'(M) \\
 \extract'(\loc{\ell}{P})                 &= \loc{\ell}{\translate^{-1}(P)}
\end{align*}

with $\translate^{-1}$ being the inverse of the function \translate{} given in the definition of \SFt.
\end{definition}

\begin{lemma}\label{lem:SF_equiv}
 For every reachable configuration, the function \SFt returns the corresponding \normalform, \IE:
 $\forall C \in \Pc\. C \equiv \SF \wedge \SF \in \Ps$.
\end{lemma}

Accordingly we show that the function \SFit returns for every \normalform a structural congruent configuration.
Note that $\SF[\SFi[\NF]] = \NF$ and $\SFi[\SF] \equiv C$.

\begin{lemma}\label{lem:SFi_equiv}
 For every \normalform \NF, $\SFi[\NF]$ returns a structural congruent configuration, \IE:
 $\forall \NF \in \Ps\. \NF \equiv \SFi[\NF]  \wedge \SFi[\NF] \in \Pr$
\end{lemma}

In the following we show that the operator $>^*$ returns a fixed point,
which is a necessary condition for the uniqueness of \standardformreps.


\begin{lemma}[Confluence]\label{lem:confluence}
 The operator $>^*$ returns a fixed point, \IE:
\[
\forall C_1, C_2 \in \Pc\. C_1 > C_2 \text{ implies } \exists! C' \in \Pc\. C_1 >^* C' \wedge C_2 >^* C'
\]
\end{lemma}

Based on the lemma above we conclude that \SF returns a unique \normalform for every reachable configuration.

\begin{lemma}\label{lem:sf_equality}
Let $C$ be a reachable configuration with $\SF$ as the corresponding \normalform.
For every other configuration $C'$ with $C \equiv C'$ the following holds: $\SF = \SF[C']$.
\end{lemma}

Finally we show that the definition of \standardformreps is unambiguous.

\begin{lemma}
 Let $C \in \Pr$ be a reachable configuration with the \normalform \NF, then $\SF = \NF$.
\end{lemma}

Now we can proof that for every configuration $C$ that has a \normalform and a step $C$ to $C'$ then there also exists a \normalform for $C'$, hence $C$ and $C'$ have \standardformreps.

\begin{theorem}\label{thm:main_theorem}
  If $\configuration \redu{\tau} \configuration'$ and $\configuration$ has a
  \normalform,
  then $\configuration'$ also has a
  \normalform.
\end{theorem}


\section{A Semantics for \StandardFormreps}
\label{sec:semant-stand-forms}

\todo[inline]{was ist denn das gute an der semantik?}

Building a Semantics for \StandardFormreps we can exploit the \StandardFormrep being a global state of the algorithm.
So we can directly derive a Semantics on \StandardFormreps out of the algorithm itself---by doing so we can ignore the actual implementation.
The only obstacle is to make sure the resulting \StandardFormrep is unique.
As  shown in Section~\ref{sec:standard-forms} the \StandardFormrep is a reordered version of the maximally evaluated calculus term.
Thus we have to keep that in mind while creating the rules.

For example in Phase~1 of the Algorithm we have to distinguish between 3 cases:
\begin{inparaenum}[(i)]
\item transition to the next Round,
\item transition to the next Phase,
\item else.
\end{inparaenum}
Additionally we have to distinguish between message reception and suspicion of the other process.
The rules that model suspicion can directly be derived from the rules that model message reception because we have to handle received values differently.

The next three rules describe the behavior of a receiving process in Phase~1 ($(q,r) \in \picol$) with the sending process not being suspected ($\lv{p}$).
We distinguish between three cases according to Line~\ref{impl:p1_if} and Line~\ref{impl:c1_if}:
\begin{compactenum}[SR1:]
  \setcounter{enumi}{2}
 \item The receiving agent is in the last round of Phase~1 receiving a message from the last agent ($p = \n \wedge r = \n-1$). 
   \setcounter{enumi}{1}
 \item The receiving agent is not in the last round of Phase~1 receiving a message from the last agent ($ p = \n \wedge r < \n-1$).
   \setcounter{enumi}{0}
 \item Else ($p < \n \wedge r < \n$).
\end{compactenum}

\begingroup
\setlength{\parindent}{0pt}
$
\exists p,q \in \P\. \lv{p} \wedge \lv{q}\. \exists r \in \N\. p < \n \wedge r < \n\. \semrule\label{rule:1} \\
(p, q, r) \in \piout \wedge
(q,r) \in \picol \wedge \Ipi[q] = p. \\
\begin{array}{crclr}
\TAB   & \pioutnew  &=& \piout \res \set{(p, q, r)} & \textup{remove the message received} \\
\wedge & \Ipinew[q] &=& \Ipi[q] + 1 & \textup{receive from the next process} \\
\wedge & \Mpinew[q] &=& \Mpi[q] \cup \set{(\comvec_{p,r}, r, p)} & \hspace{7.4em} \textup{add the received information to local messages} \\
\end{array}
$
\endgroup

\bigskip

\begingroup
\setlength{\parindent}{0pt}
$
\exists p,q \in \P\. \lv{p} \wedge \lv{q}\. \exists r \in \N\. p = \n \wedge r < \n-1. \semrule\label{rule:2} \\
(p, q, r) \in \piout \wedge
(q,r) \in \picol \wedge \Ipi[q] = p. \\
\begin{array}{crclr}
\TAB   & \pioutnew          &=& \bigcup_{j=1}^{n} \set{(q,j,r+1)} \cup (\piout \res \set{(p, q, r)}) & \textup{additionally send messages to all other proc} \\
\wedge & \Ipinew[q]         &=&  1 & \textup{receive from the first process} \\
\wedge & \Mpinew[q]         &=& \Mpi[q] \cup \set{(\comvec_{p,r}, r, p)} \\
\wedge & \Vpinew[q]         &=& \updatek(r, \Mpinew[q], \Vpi[q]) & \textup{update knowledge} \\
\wedge & \comvecnew_{q,r+1} &=& \updater(r, \Mpinew[q], \Vpi[q]) & \textup{update newly received knowledge} \\
\wedge & \picolnew          &=& \left( \picol \res \set{(q,r)} \right) \cup \set{(q, r+1)} & \textup{collect messages of the next round} \\
\end{array}
$
\endgroup

\bigskip

The change to $\Ipinew[q]$ may be omitted in the following rule because no according triple is added to the set $\pioutnew$.
Because $\picolnew = \picol \res \set{(q,r)}$ we may also omit the change to $\Ipinew[q]$.
Both changes have no effect on the derived \standardformrep.

\begingroup
\setlength{\parindent}{0pt}
$
\exists p,q \in \P\. \lv{p} \wedge \lv{q}\. \exists r \in \N\. p = \n \wedge r = \n-1. \semrule\label{rule:3}\\
(p, q, r) \in \piout \wedge
(q,r) \in \picol \wedge \Ipi[q] = p.\\
\begin{array}{crclr}
\TAB   & \pioutnew          &=& \piout \res \set{(p, q, r)} \\
\wedge & \Ipinew[q]         &=& \Ipi[q] + 1 \\
\wedge & \piioutnew         &=& \bigcup_{j=1}^{n} \set{(q,j)} \cup \piiout & \textup{send messages in Phase 2} \\
\wedge & \Mpinew[q]         &=& \Mpi[q] \cup \set{(\comvec_{p,r}, r, p)} \\
\wedge & \Vpinew[q]         &=& \updatek(r, \Mpinew[q], \Vpi[q]) \\
\wedge & \comvecnew_{q,r+1} &=& \updater(r, \Mpinew[q], \Vpi[q]) \\
\wedge & \picolnew          &=& \picol \res \set{(q,r)} & \hspace{10.5em} \textup{stop receiving messages in Phase 1} \\
\wedge & \piicolnew         &=& \piicol \cup \set{q} & \textup{receives messages in Phase 2}\\
\wedge & \Ipiinew[q]        &=& 1 \\
\wedge & \Vpiinew[q]        &=& \Vpinew[q] & \textup{transfer knowledge to Phase 2} \\
\wedge & \Mpiinew[q]        &=& \Mpinew[q] & \textup{transfer knowledge to Phase 2} \\
\end{array}
$
\endgroup

\smallskip

The Rules \setcounter{rulecount}{4}\therulecount\label{rule:4} to \setcounter{rulecount}{6}\therulecount\label{rule:6} are nearly the same as Rules~\ref{rule:1} to \ref{rule:3}.
They model suspicion of the sending process, thus the only differences in the Rules are:
\begin{inparaenum}[(i)]
\item the message in transit does not get removed,
\item the value that should be received gets replaced by $\bot$.
\end{inparaenum}

\bigskip


The next rule applies if some agent has crashed, \IE the process and all of its messages gets removed from the system.

\begingroup
\setlength{\parindent}{0pt}
$
\exists p \in \P\. p \not= \ti\ \wedge \lv{p}. \semrule\label{rule:7}\\
\begin{array}{crcl}
\TAB   & (\widehat{\L}, \widehat{n}) &=& (\L \res \set{p}, n-1) \\
\wedge & \pioutnew                   &=& \piout \res \makeset{(p, i, r)}{\exists i \in \P,r \in \N\. (p,i,r) \in \piout} \\
\wedge & \piioutnew                  &=& \piiout \res \makeset{(p,i)}{\exists i \in \P\.(p,i) \in \piiout} \\
\wedge & \piiioutnew                 &=& \piiiout \res \set{p} \\
\wedge & \picolnew                   &=& \picol \res \makeset{(p,r)}{\exists r \in \N\. (p,r) \in \picol} \\
\wedge & \piicolnew                  &=& \piicol \res \set{p} \\
\end{array}
$
\endgroup

\subsection{Correctness}
\label{sec:correctness}

\todo[inline]{mehr fokus was das eigentlich bringt, also was ist toll an 1-zu-1}

\todo[inline]{After having shown sound and corr we can freely choose either the standard form semantics or the calculus semantics to prove the invariants. Thus we can freely switch between state based and action based reasoning whenever needed.}

To obtain a 1-1-correspondence, we prove our semantics to be equivalent to the calculus semantics restricted to \normalforms, see \figref{fig:soundness} and \ref{fig:completeness}.
Therefore we show two main properties which we call \emph{soundness} and \emph{completeness}.
\begin{compactdesc}
 \item[Soundness.] We call our semantics to be sound, if whenever a \normalform \NF can do a step to $\NF'$, the according configuration $C$ can do a step to $C'$ such that $\NF'$ is the \normalform to $C'$.
 \item[Completeness.] We call our semantics to be complete, if for every step of a configuration $C$ to $C'$,  the according \normalform \NF can do a step to $\NF'$ such that $\NF'$ is the according \normalform to $C'$.
\end{compactdesc}
%
By showing these properties we are able to solely use the semantics of \standardformreps to show properties of the implementation itself.

Being able to focus on \standardformreps enables us to tame the state space, because by Definition~\ref{def:standard_form} a \normalform is structural congruent to a configuration and by Lemma~\ref{lem:sf_equality} every two structurally equivalent configurations have the same \normalform.
Hence a \standardformrep is a representative of an equivalence class of configurations.

   

The semantics is sound regarding to the calculus if every step in the \standardformrep semantics can be emulated by the calculus semantics, see \figref{fig:soundness}.

\begin{theorem}[Soundness]\label{thm:soundness}
$\forall \NF, \NF' \in \Ps\. \NF \redu{} \NF' \text{ implies } \exists C, C' \in \Pc\. \SFi[\NF] = C \wedge \SFi[\NF'] = C' \wedge C \redu{\tau} C'$.
\end{theorem}

To show soundness we have to provide a proof tree using the calculus semantics for each rule of the \standardformrep semantics.
By proving soundness we observe that we obtain 7 proof trees, two for each phase and one for process crashes.

\begin{figure}[tbp]
  \begin{floatrow}
    \ffigbox{\begin{tikzpicture}[node distance=2.5cm,auto,>=latex']
        \node (c)                                  {$\NF$};
        \node (c')  [below of=c]                   {$\NF'$};
        \node (sf)  [right of=c,node distance=4cm] {$\SFi[\NF]$};
        \node (sf') [below of=sf]                  {$\SFi[\NF']$};
        
        \path[->] (c) edge node [swap] {$\boldsymbol\forall$} (c');
        \path[->, dashed] (sf) edge node [swap] {$\boldsymbol\exists$} (sf');
        \draw [->,decorate,
        decoration={snake,amplitude=.4mm,segment length=2mm,post length=1mm}]
        (c) -- (sf);
        \draw [->,decorate,
        decoration={snake,amplitude=.4mm,segment length=2mm,post length=1mm}]
        (c') -- (sf');
      \end{tikzpicture}}
    {\caption{Soundness}\label{fig:soundness}}
    \ffigbox{\begin{tikzpicture}[node distance=2.5cm,auto,>=latex']
        \node (c)                                  {$\SF$};
        \node (c')  [below of=c]                   {$\SF[C']$};
        \node (sf)  [right of=c,node distance=4cm] {$C$};
        \node (sf') [below of=sf]                  {$C'$};
        
        \path[->, dashed] (c) edge node [swap] {$\boldsymbol\exists$} (c');
        \path[->] (sf) edge node [swap] {$\boldsymbol\forall$} (sf');
        \draw [->,decorate,
        decoration={snake,amplitude=.4mm,segment length=2mm,post length=1mm}]
        (sf) -- (c);
        \draw [->,decorate,
        decoration={snake,amplitude=.4mm,segment length=2mm,post length=1mm}]
        (sf') -- (c');
      \end{tikzpicture}}
    {\caption{Completeness}\label{fig:completeness}}
  \end{floatrow}
\end{figure}

To prove completeness we have to show that every transition being possible in the calculus beginning from the initial setting can be simulated by our \standardformrep semantics, see \figref{fig:completeness}.
For simplicity in the following we always assume $\live$ for the sending and receiving agent in a communication step.
In the case of suspicion we assume $\live$ for the receiving agent and neither $\live$ nor $\neg\live$ for the sending agent.
Note that neither the wrapper nor the trusted immortal can crash.

\begin{theorem}[Completeness]\label{thm:completeness}
$\forall C,C' \in \Pc\. C \redu{\tau} C' \text{ implies } \exists \NF, \NF' \in \Ps\. \NF = \SF \wedge \NF' = \SF[C'] \wedge \NF \redu{} \NF'$.
\end{theorem}

To prove this Theorem we use Lemmata \ref{lem:normalform_for_reachables}, \ref{lem:SF_equiv}, and \ref{lem:sf_equality} to be able to only observe \normalforms instead of every reachable configuration.
Analyzing the \normalform we observe that the only possible steps are either communication or process crashes.
For each of these cases we show that there is a corresponding rule in the \standardformrep semantics, \EG communication in Phase~1 corresponds to Rules~\ref{rule:1} to \ref{rule:3}.
We show by case differentiation that for each communication step in Phase~1 exactly one of these rules can be applied to obtain the correct \standardformrep.
The proof for the other cases is similar.

Soundness and correctness allows us to freely switch between both semantics.
Thus we can now use state based reasoning techniques like invariants as well as action based techniques like bisimulation at the same time.


\section{Conclusion}
\label{sec:conclusion}

The main ideas have been explained in the Introduction. We may finally
emphasize that we could have started the whole verification exercise from
scratch using a state machine approach, for example using the ASM formalism
or also TLA+.  Actually, we have done such work ourselves, as in
\cite{nestmann07} as the basis for fully formal verification \cite{kuefner.etal-TCS12} within the
Isabelle proof checking environment, but we are still convinced that the initial design of
the algorithm using language primitives to ``code'' the local behaviors
for the individual processes is of good value.  After all, it is then very
easy for programmers to implement the algorithm on that basis.  In
comparison, this would be rather indirect when starting from a state
machine approach.


\bibliographystyle{eptcs}
\bibliography{references}



\end{document}